\documentclass[epjCONF]{svjour}
\usepackage[varg]{txfonts} 
\usepackage[numaddn=\pi,per=fraction]{siunitx}
\usepackage{subfigure}
\usepackage{booktabs, tabularx}

\usepackage{multicol}
\usepackage[switch]{lineno}
\usepackage{graphicx}  

\session-title{UHECR2012 Symposium}

\begin{document}
\title{ Disentangling the Air Shower Components Using Scintillation and
  Water Cherenkov Detectors  }

\author{Javier G. Gonzalez \and Markus Roth \and Ralph Engel }
\institute{Institute for Nuclear Physics, Karlsruhe Institute of Technology (KIT)}


\abstract{
We consider a ground array of scintillation and
water Cherenkov detectors with the purpose of determining the muon
content of air showers. The different response characteristics of
these two types of detectors to the components of the air shower
provide a way to infer their relative contributions. We use a detailed
simulation to estimate the impact of parameters, such as
scintillation detector size, in the determination of the size of the
muon component.
}

\maketitle
\section{Introduction}

The measurement of the mass composition of ultra-high energy cosmic
rays is one of the keys that can help us elucidate their origin. Using
a surface detector array to disentangle the contributions to the
detector signal from the different components of an air shower is a
way to accomplish this.

Up to energies around \SI{e15}{eV}, the Galaxy is believed to be the
source of cosmic rays. Several acceleration mechanisms are certainly
at play but it is widely expected that the dominant one is first order
Fermi acceleration at the vicinity of supernova remnant shock
waves. These Galactic accelerators should theoretically become
inefficient between \SI{e15}{eV} and \SI{e18}{eV}. The KASCADE
experiment has measured the energy spectra for different mass groups
in this energy range and found that there is a steepening of the
individual spectra at an energy that increases with the cosmic ray mass
\cite{Apel:2008cd}. As a result, the mass composition becomes
progressively heavy. It is also thought that extra-galactic sources
can start to contribute to the total cosmic ray flux at energies above
\SI{4e17}{eV}. The onset of such an extra-galactic component would
probably produce another change in composition.  The $X_{\text{max}}$
measurements from the HiRes-MIA experiment have been interpreted as a
change in composition, from heavy to light, starting at \SI{4e17}{eV}
and becoming proton-dominated at \SI{1.6e18}{eV}
\cite{AbuZayyad:2000ay,Abbasi:2009nf}.
Both HiRes and the Pierre Auger Observatory has measured a suppression of the flux of
cosmic rays at the highest energies \cite{Abbasi:2005ni,Abraham:2010mj} and
the $X_{\text{max}}$ measurements hint at a light or mixed composition
that becomes heavier beyond \SI{2e18}{eV} \cite{Abraham:2010yv}.



Roughly speaking, the techniques for inferring the mass composition of
cosmic rays can be split in two categories, depending on whether they exploit the
sensitivity to the depth of shower maximum ($X_{\text{max}}$) or to the ratio of the
muon and electromagnetic components of the air shower
\cite{LetessierSelvon:2011dy}. Direct measurements of the fluorescence
emission fall in the first category, and so do the various
measurements of the Cherenkov light produced by air showers. Most ground-based
detector observables depend one way or another on the number of
muons in the air shower. However, the arrival time profile of shower
particles has been used as an observable mostly sensitive to
$X_{\text{max}}$, in particular the so-called \textit{rise-time}, the time it
takes for the signal to rise from 10\% to 50\% of the integrated
signal \cite{Wahlberg:2009zz}.
The measurement of the number of muons and electrons in an air shower
can be done directly, for example, the way it was done with the
KASCADE detector \cite{Antoni:2003gd}.

It is important that
we disentangle the contributions from the different components of the
air shower, as this relates to the primary mass as well as possible
systematic uncertainties arising from the use of Monte Carlo hadronic
interaction generators.

The Pierre Auger Observatory is developing a series of enhancements
that aim at measuring showers in the energy range between \SI{e17}{eV} and \SI{e19}{eV}
\cite{amiga_2011,heat}. In particular, the objective of the AMIGA
enhancement \cite{amiga_2011} is the measurement of the muon component of the air
showers using scintillators shielded by several meters of soil. In the
same spirit, we are considering a \textit{combined} surface array,
consisting of two super-imposed ground arrays, a Water Cherenkov
Detector (WCD) array and a scintillation detector array. The purpose
of the scintillation detectors is to increase the sensitivity to the
electromagnetic component of air showers.

In this article we consider the possibility of studying extensive air
showers induced by cosmic rays using a combined detector consisting of
water Cherenkov and scintillation detectors. In order to do this, we
have developed a detailed simulation and reconstruction chain whose characteristics
will be briefly described in Section \ref{section:simulation}. We will then
look at the general features that allow us to gain sensitivity to the
mass composition of cosmic ray primaries at energies around
\SI{e18}{eV} and conclude, in Section \ref{sect:e19.5_eV}, by
considering the possibility of determining the contribution from the
muon component of a shower from a pair of scintillation and WCD
detectors.

\section{Studying the Characteristics of a Combined Surface Detector}
\label{section:simulation}

The detector setup we are considering consists of an array of WCDs and an
array of scintillators, both covering the same area. Each WCD is like
a typical Pierre Auger detector. That is: it is made of a 12
ton cylindrical water tank with \SI{10}{m^2} top surface area and the
light is collected by three \SI{9}{inch} photomultipliers placed at the
top of the tank facing down. The light collected on the PMTs is then
digitized by a 40 MHz Flash ADC.

The array of scintillation detector stations is arranged in a regular
grid. We have considered different grid configurations and these will
be specified in Sections \ref{sect:e18_eV} and
\ref{sect:e19.5_eV}. Each scintillation detector station is made of
\SI{3}{cm} thick plastic scintillator tiles. In order to enhance the
signal from gamma-rays in air showers, we have studied the effect of
adding a certain amount of lead on top of the scintillators. For
conversion of around 80\% of the high energy gamma-rays one normally
needs a shielding of about 2 radiation lengths. One radiation length
corresponds to \SI{0.56}{cm} of lead.

In order to study such a combined detector, we have implemented a
simulation and reconstruction chain based on the Pierre Auger
Observatory offline framework \cite{Argiro:2007qg}. All showers were
generated using CORSIKA \cite{corsika}, with QGSJET II for high energy hadronic interaction
simulations \cite{Ostapchenko:2010vb}. The specific zenith angles and
energies studied will be mentioned in Sections \ref{sect:e18_eV} and
\ref{sect:e19.5_eV}. The simulation of the interactions of the shower
particles with the detector is done using the Geant4 package
\cite{geant4,geant4_2}. The scintillation efficiencies used in the
simulation correspond to the specifications for Bicron's BD-416
scintillators: Polyvinyl Toluene scintillators with a nominal light
yield of about \SI{e4}{photons/MeV} and a density of
\SI{1.032}{g/cm^3}. The resulting scintillation photons are sampled with a 100\% efficiency at
a frequency of \SI{100}{MHz} to produce one FADC trace per
station. The signal in each station is measured in units of
\textit{Minimum Ionizing Particle} equivalent, or \textit{MIP}, where
a MIP is given by the position of the peak of the Landau distribution
for vertical muons. We then consider only stations with signals
between 1 and \SI{2000}{MIP} in order to simulate a limited dynamical range.

The arrival direction and core
position of each event are estimated using only the WCDs. The arrival
direction is determined by fitting a spherical shower front to the
signal start times of the stations in the event. The core position is
determined by adjusting a lateral distribution function (LDF) of the
form
\begin{equation}
  \left(\frac{r}{S_{r_0}}\right)^{\beta} \left(\frac{r + 700 \text{m}}{S_{r_0} + 700 \text{m}}\right)^{ \beta + \gamma}
  \label{eq:LDF}
\end{equation}
to the total signal in the stations in the event. The $r_0$ parameter
depends on the WCD array grid spacing. It will be \SI{450}{m} when the
stations are separated by \SI{750}{m} and \SI{1000}{m} when they are
separated by \SI{1500}{m}. Correspondingly, $S_{450}$ and $S_{1000}$
are the usual energy estimators for a WCD array with these grid
spacings as they are close to the optimum distance for determining the
signal in each case \cite{Maris:2011zz,Newton:2006wy}.

\section{Measuring Composition at Energies around 10$^{18}$ eV}
\label{sect:e18_eV}

We have considered various configurations in order to estimate the
cosmic ray primary composition at energies around \SI{e18}{eV}. In a
previous contribution we showed that the addition of lead converters
on top of the scintillator stations to enhance the
contribution from photons in the shower does not increase the
sensitivity to the primary mass \cite{scint_array-icrc11}.
We have also considered various spacings between scintillator
detectors, while keeping the total array area as well as the total
collecting area constant. We have considered three
such arrangements for the scintillator array, corresponding to
three different spacings: 433, 612, and 750 \SI{}{m} regular
triangular grid. The area of the scintillator stations in these
configurations are 3.2, 6.4, and 9.7 \SI{}{m^2} respectively.

\begin{figure}[t]
  \begin{minipage}[t]{0.47\linewidth}
    \includegraphics[width=\textwidth]{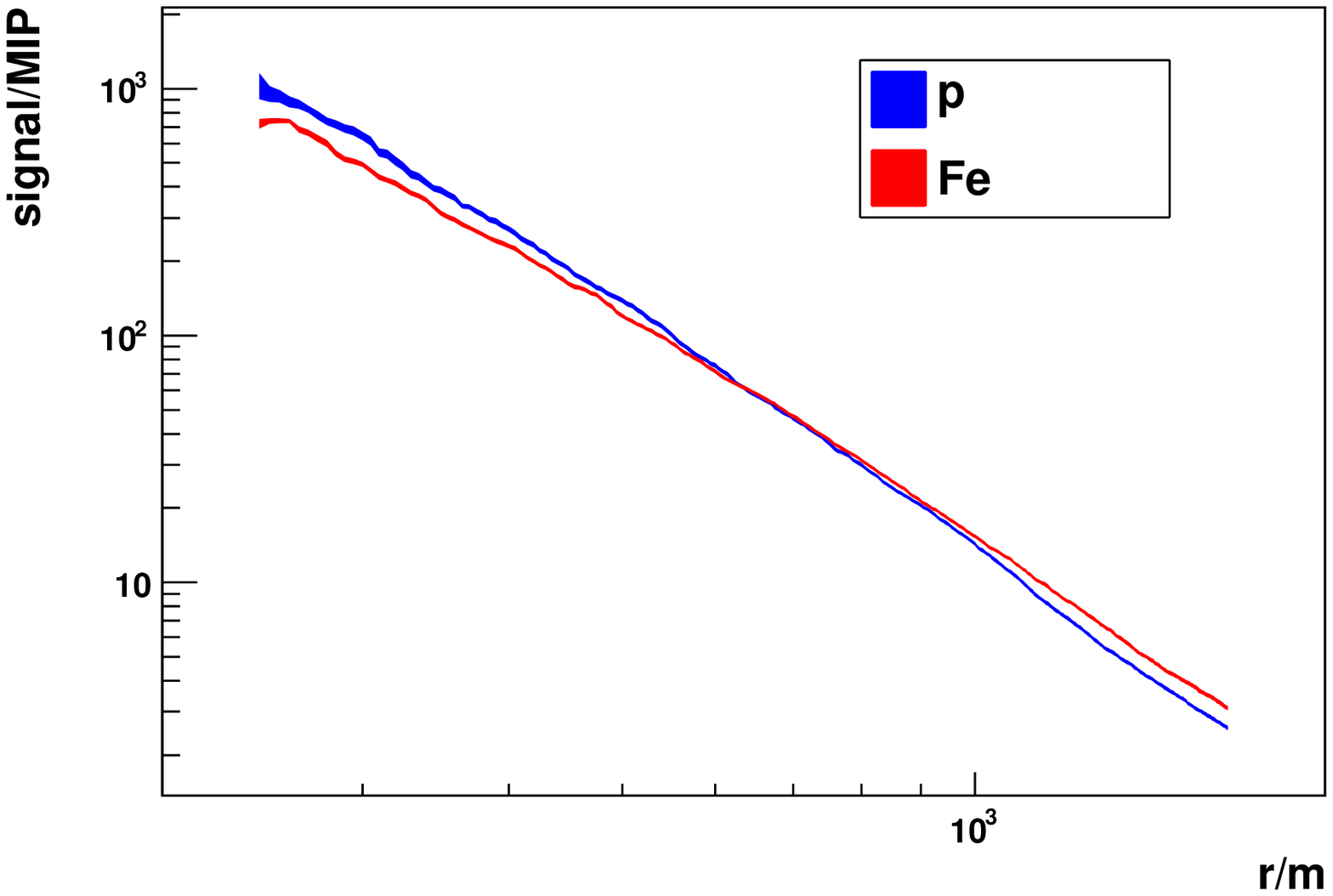}
    \caption{p-Fe LDF Comparison at \SI{38}{\degree} (E=10$^{18}$ eV).
      The lines correspond to the average over 400 showers.
    }
    \label{fig:p_fe_ldf_comparison}
  \end{minipage}
  \hspace{0.05\linewidth}
  \begin{minipage}[t]{0.47\linewidth}
    \includegraphics[width=\textwidth]{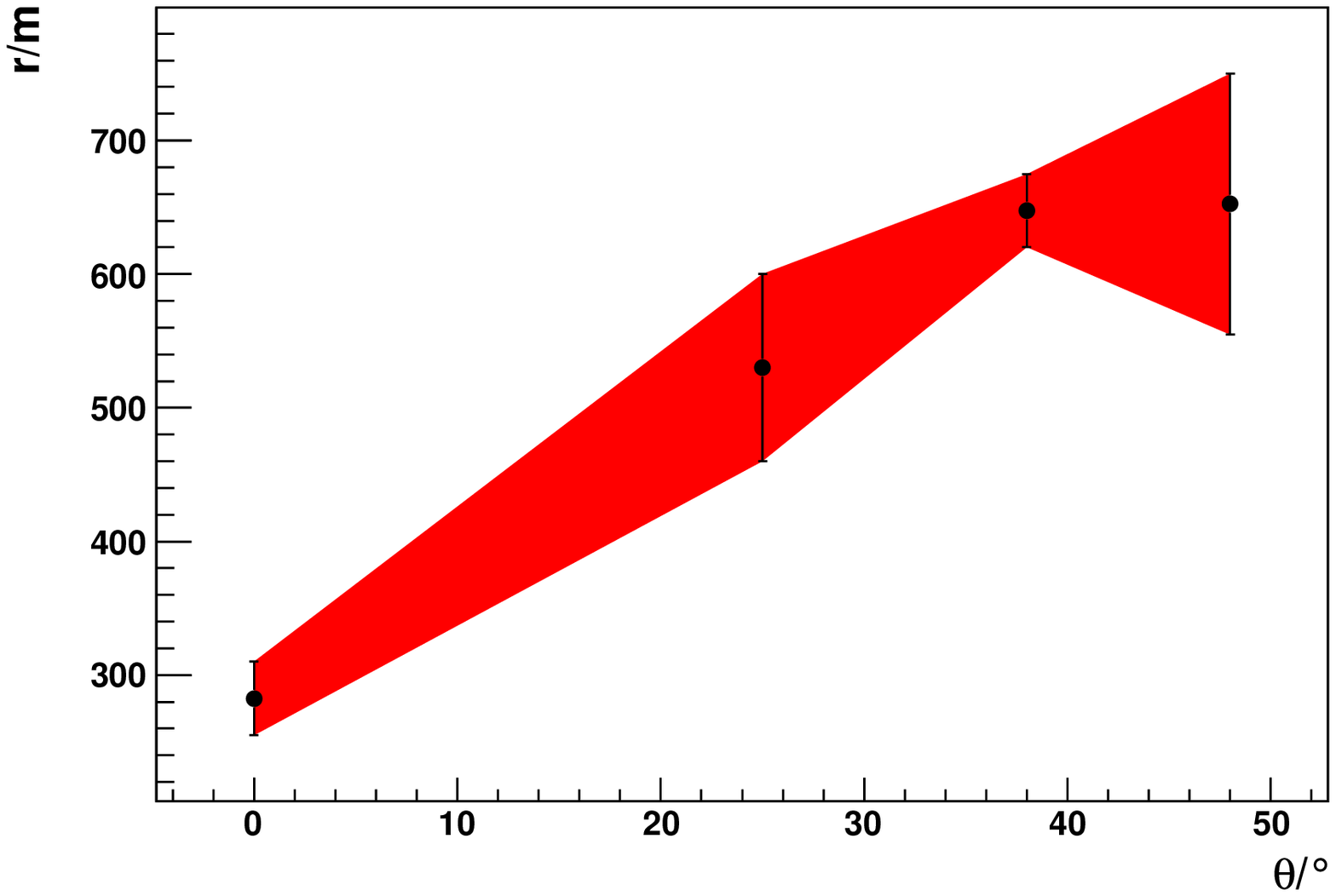}
    \caption{p-Fe LDF Crossing Points ($\theta$=\SI{38}{\degree}, E=10$^{18}$ eV)}
    \label{fig:p_fe_ldf_crossing}
  \end{minipage}
\end{figure}

One can see in Figure \ref{fig:p_fe_ldf_comparison} that the average
LDF from light primaries is steeper than that from heavier
primaries. In order to be sensitive to the mass of the primary, one
needs to sample the LDF at points away from the crossing point of the
two LDFs. This crossing point depends on energy as well as zenith
angle, as displayed in Figure \ref{fig:p_fe_ldf_crossing}. We need
therefore to measure the signal either close to or far away from the
shower axis. Since we are considering this scintillator array to be
placed around a base configuration consisting of a fixed-size WCD
array, it follows that the scintillators must probe the region close
to the axis.

\begin{figure}[t]
  \begin{minipage}[t]{0.47\linewidth}
    \includegraphics[clip=true,width=\textwidth]{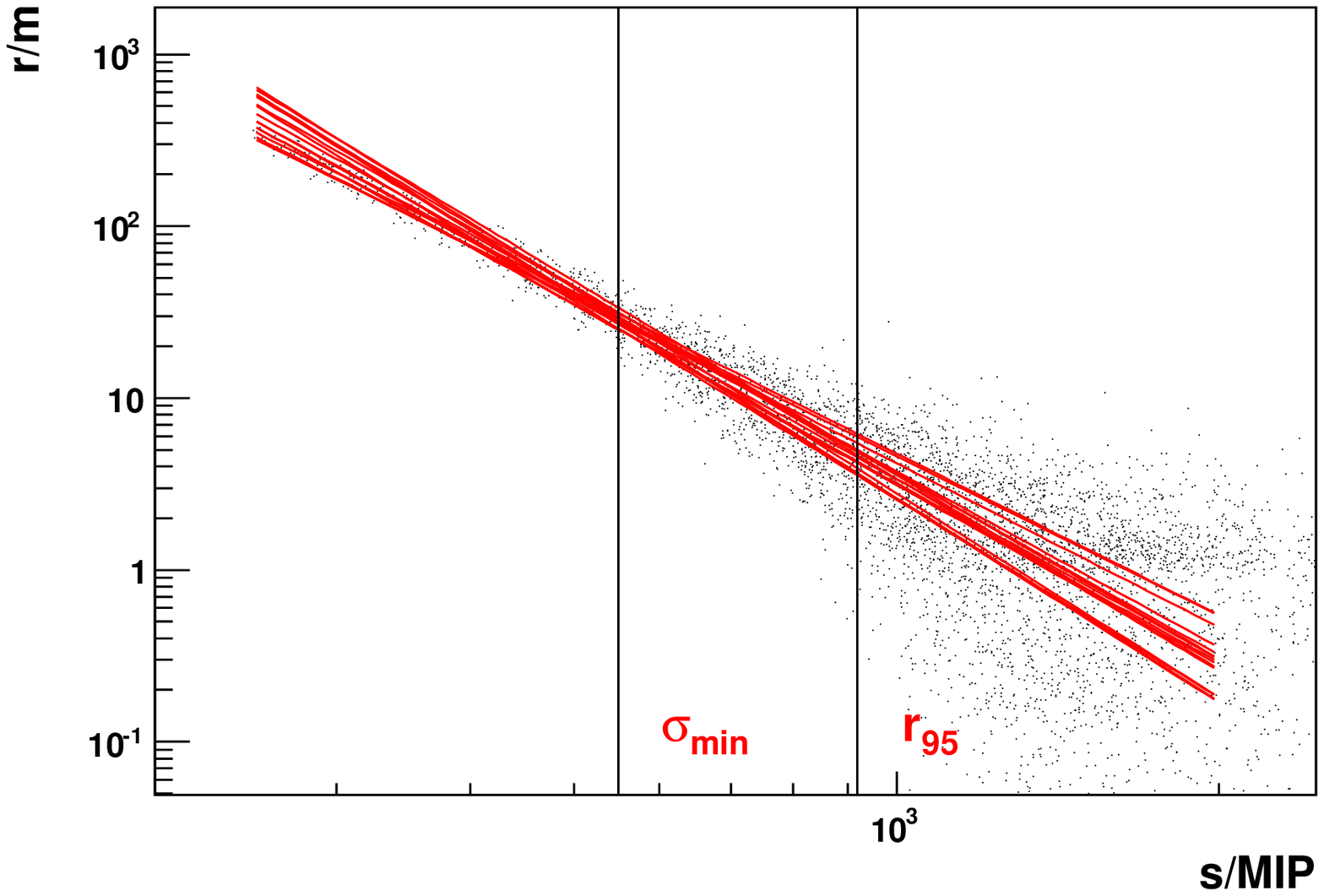}
    \caption{A set of reconstructed scintillator LDFs (proton, $\theta$=\SI{38}{\degree}, E=10$^{18}$ eV).}
    \label{fig:event_ldfs}
  \end{minipage}
  \hspace{0.05\linewidth}
  \begin{minipage}[t]{0.47\linewidth}
    \includegraphics[clip=true,width=\textwidth]{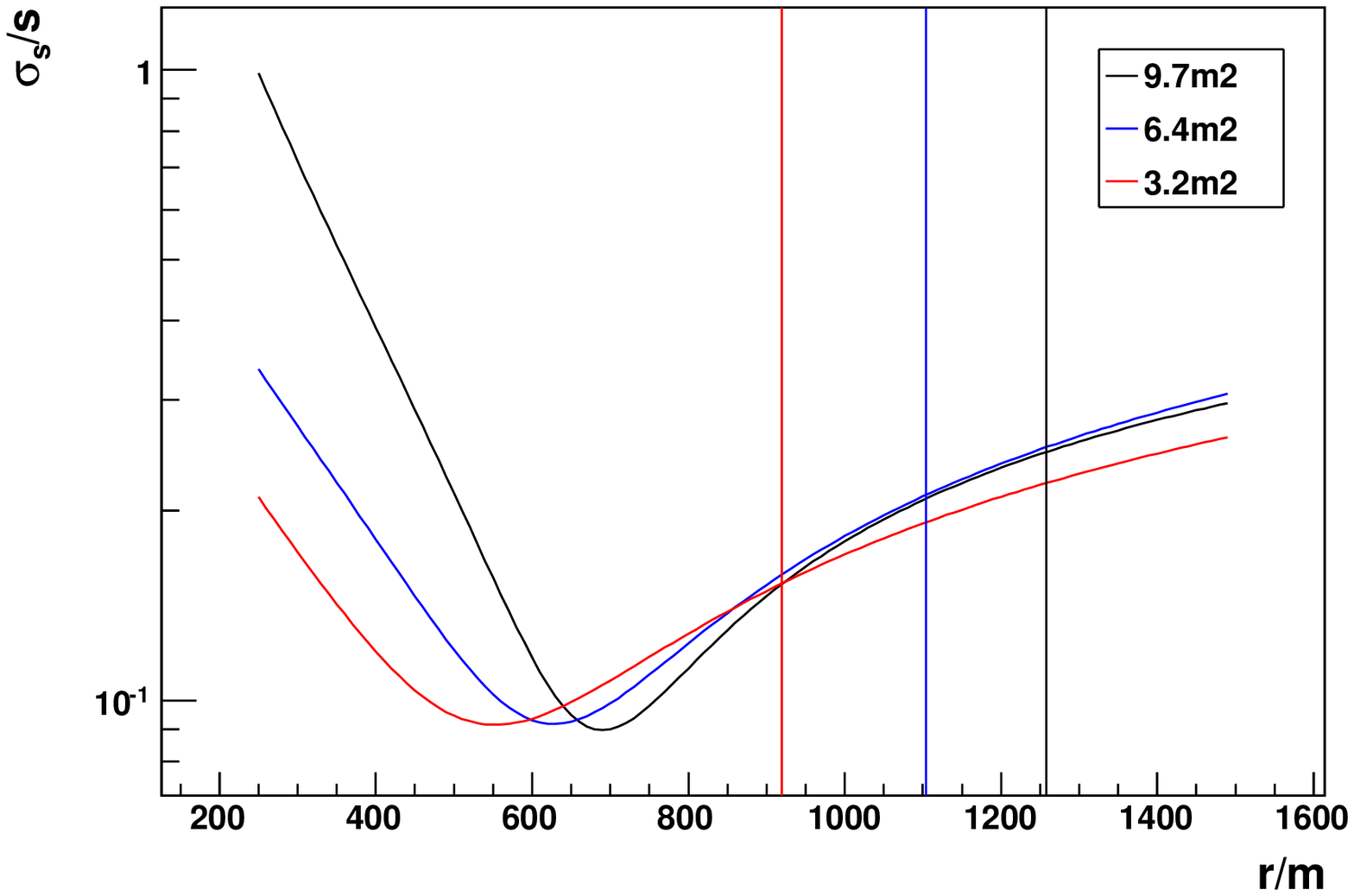}
    \caption{Relative signal fluctuations for the different array configurations considered. Same parameters as in fig. \ref{fig:event_ldfs}.}
    \label{fig:signal_fluctuations}
  \end{minipage}
\end{figure}

The signals from the scintillation detectors are used to estimate an
LDF on an event-by-event basis using a function of the form
\begin{equation}
  S_{\text{sci}}(r)=S_{r_0}\,\left(\frac{r}{r_0}\right)^{-\beta}.
  \label{eq:LDF2}
\end{equation}
A sample of a few reconstructed LDFs is depicted in Figure
\ref{fig:event_ldfs}. In this figure one can see that there will be an
optimum distance from the shower axis to measure the scintillator
signal. The optimum radial distance will be the one where the spread
of the reconstructed signal relative to the total reconstructed signal
is minimal. This is shown in Figure \ref{fig:signal_fluctuations},
where we display the relative signal fluctuations for the different
configurations used. For reference, in both figures, we display the
radius at which the local trigger is 95\% efficient, $r_{95}$. At this
point it becomes clear that trading off individual detector size for a
denser array will increase the composition sensitivity, since the
optimum distance will be displaced closer to the axis.

\begin{figure}[t]
  \begin{minipage}[t]{0.47\linewidth}
    \subfigure[\footnotesize $\theta$=0$^{\circ}$]{
      \includegraphics[width=\linewidth]{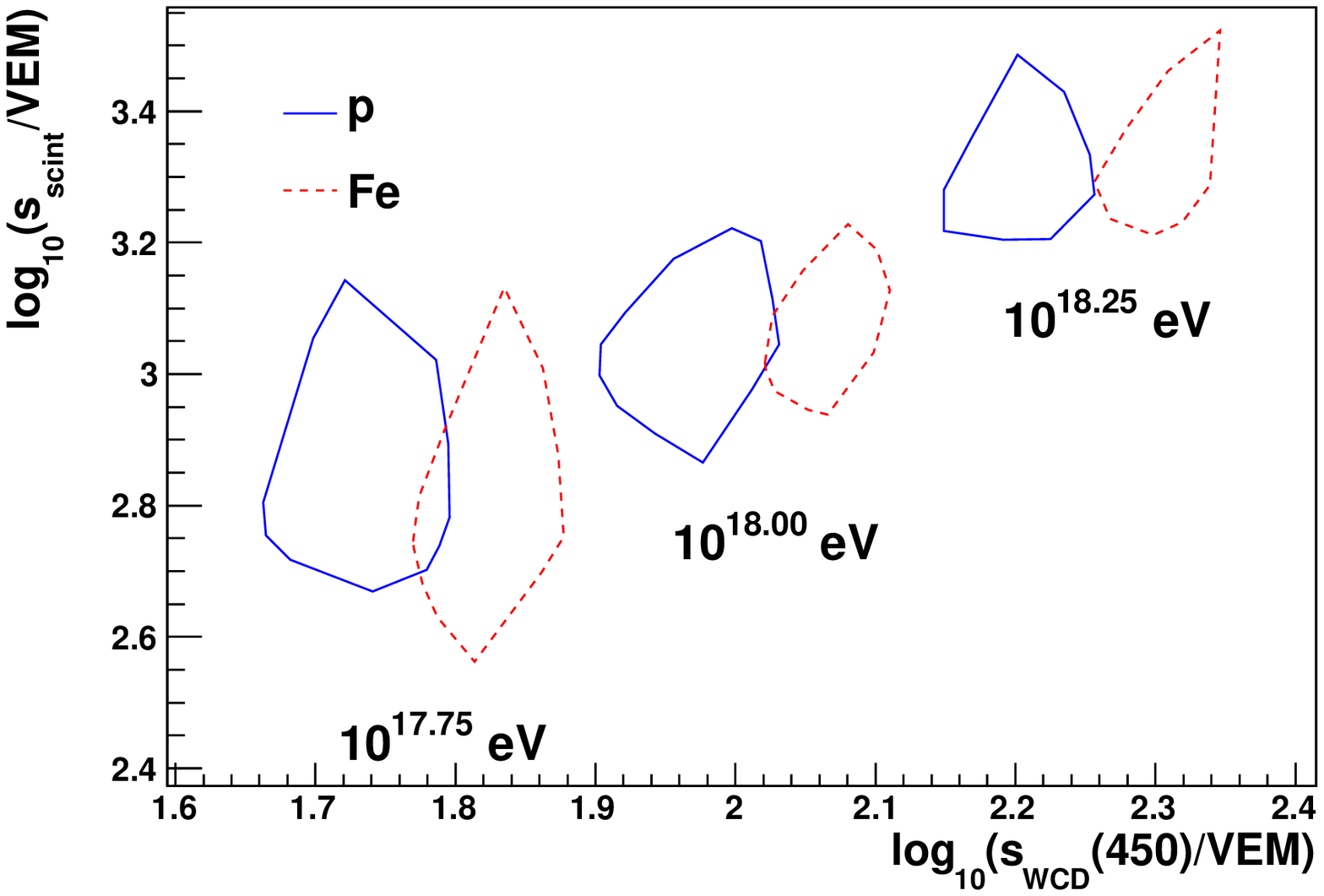}\\
    }
    \subfigure[\footnotesize $\theta$=28$^{\circ}$]{
      \includegraphics[width=\linewidth]{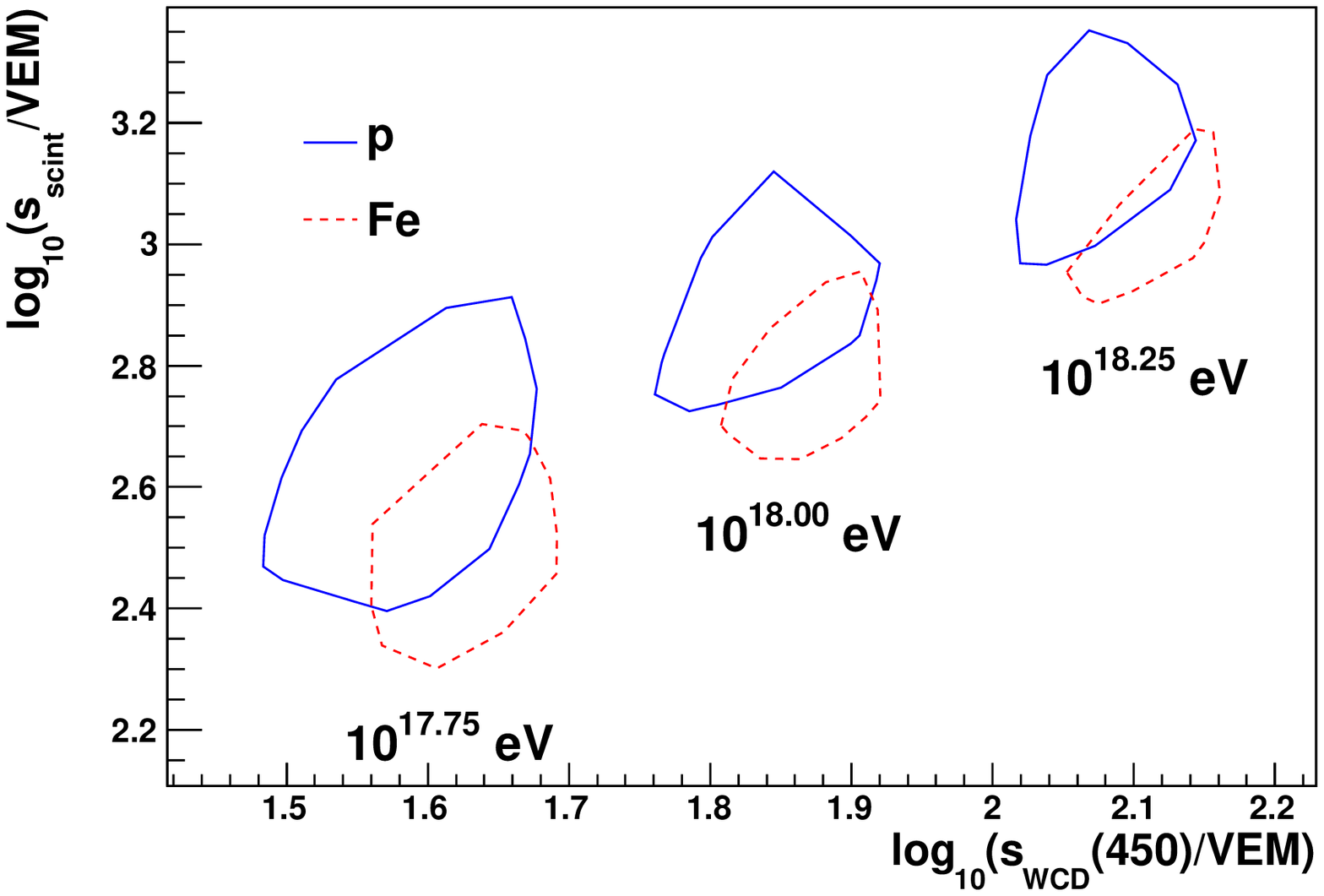}
    }
  \end{minipage}
  \hspace{0.05\linewidth}
  \begin{minipage}[t]{0.47\linewidth}
    \subfigure[\footnotesize $\theta$=25$^{\circ}$]{
      \includegraphics[width=\linewidth]{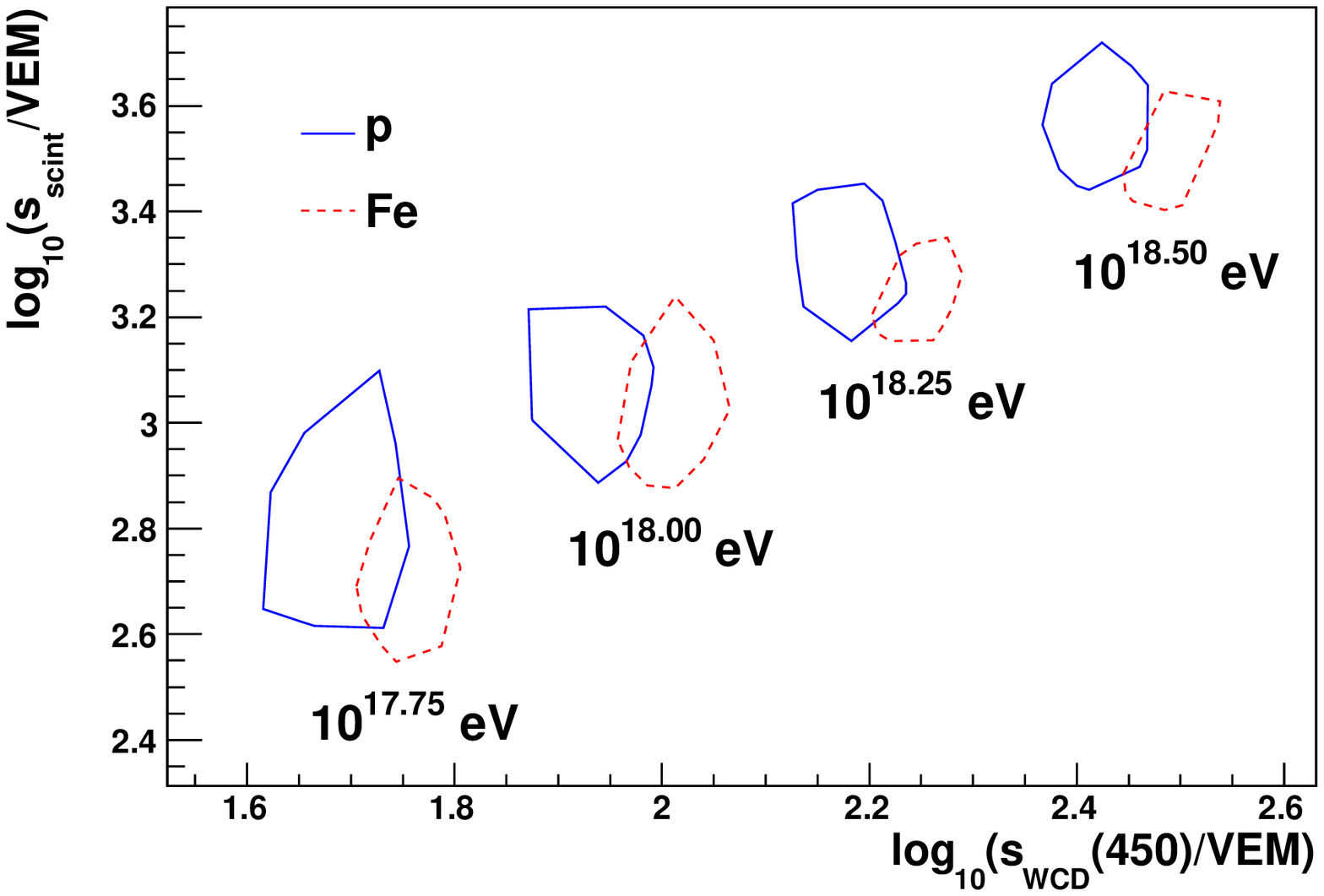}\\
    }
    \subfigure[\footnotesize $\theta$=48$^{\circ}$]{
      \includegraphics[width=\linewidth]{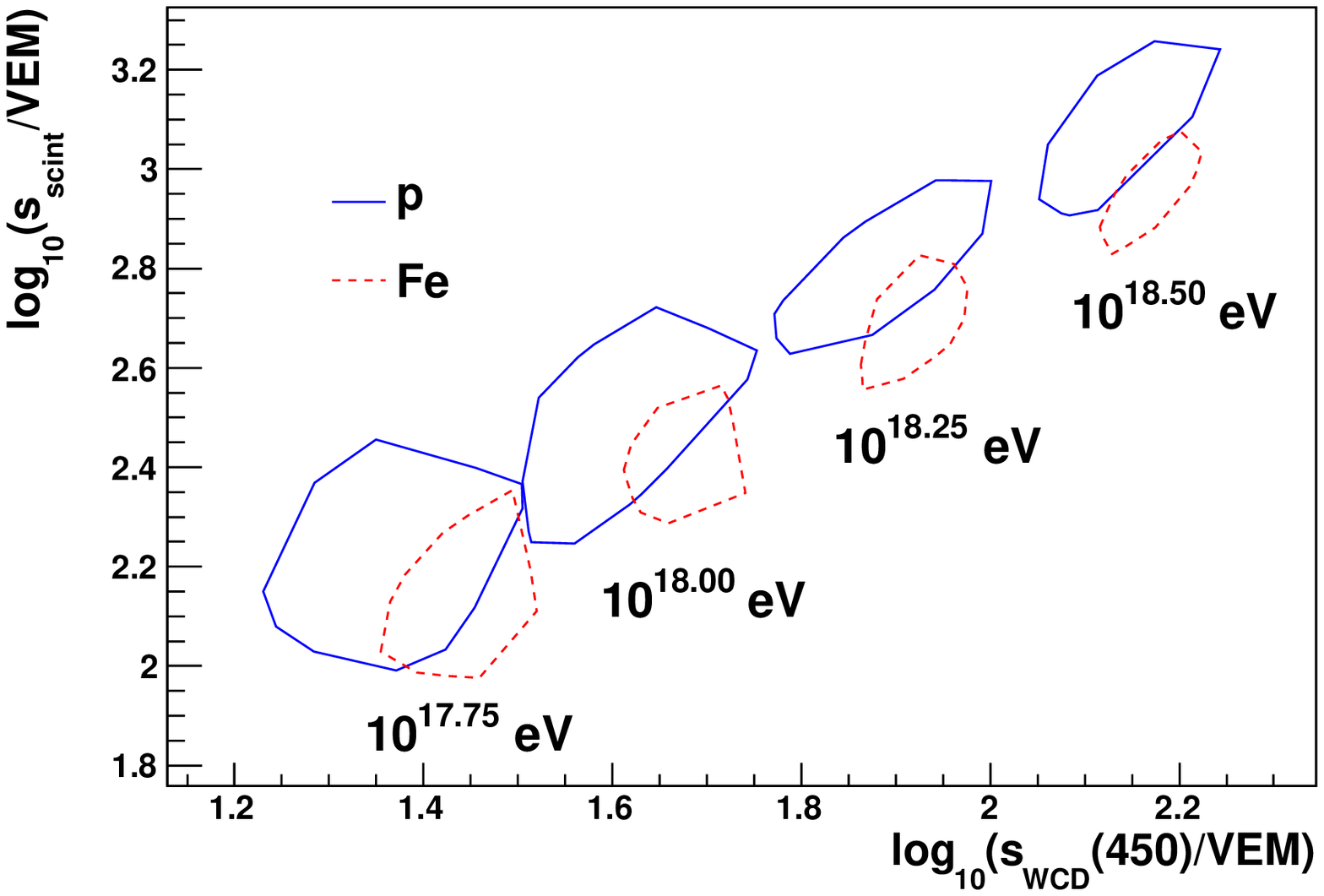}
    }
  \end{minipage}
  \caption{68\% Contours of the $\log_{10}(S_{\text{sci 400m}})$ − $\log_{10}(S_{\text{WCD}})$.
}
  \label{fig:blobs}
\end{figure}

In a similar way, the optimum distance for composition sensitivity is
found by minimizing the signal fluctuations in relation to the average
separation of the proton and iron LDFs. For this reason one can choose
slightly smaller distances to the shower axis. We can then correlate
the recorded scintillator signal at a specified distance with the WCD
at \SI{450}{m}. This correlation, with the scintillator signal
measured at 400 m, can be seen in Figure \ref{fig:blobs}.  From the
$\log_{10}(S_{\text{sci}})$ − $\log_{10}(S_{\text{WCD}})$ correlation
it is possible to provide an estimator of the primary mass
composition.

\section{Estimating the Muon Signal At the Highest Energies}
\label{sect:e19.5_eV}

In the previous Section we discussed how to use an array of
scintillators to measure the electromagnetic (EM) component of air
showers and found that decreasing the spacing between the
scintillators increases the sensitivity to the primary
mass composition. The LDF of the electromagnetic component is steeper,
therefore we need a denser array in order to reconstruct it.

We now turn to the other extreme. We will consider how well we can
determine the contributions from the EM and muon
components in a single pair of WCD/scintillator stations, one next to
the other.  At the highest energies, the electromagnetic LDF will
extend to larger distances and, while it will then not be possible to
reconstruct it using an array of scintillators separated the same
distance as the WCD detectors, a significant fraction of events will
contain at least a pair of WCD/scintillator stations with which to
determine the contribution from the EM and muon components.

\begin{figure}[t]
  \begin{minipage}[t]{0.47\linewidth}
  \begin{center}
    \includegraphics[width=\linewidth]{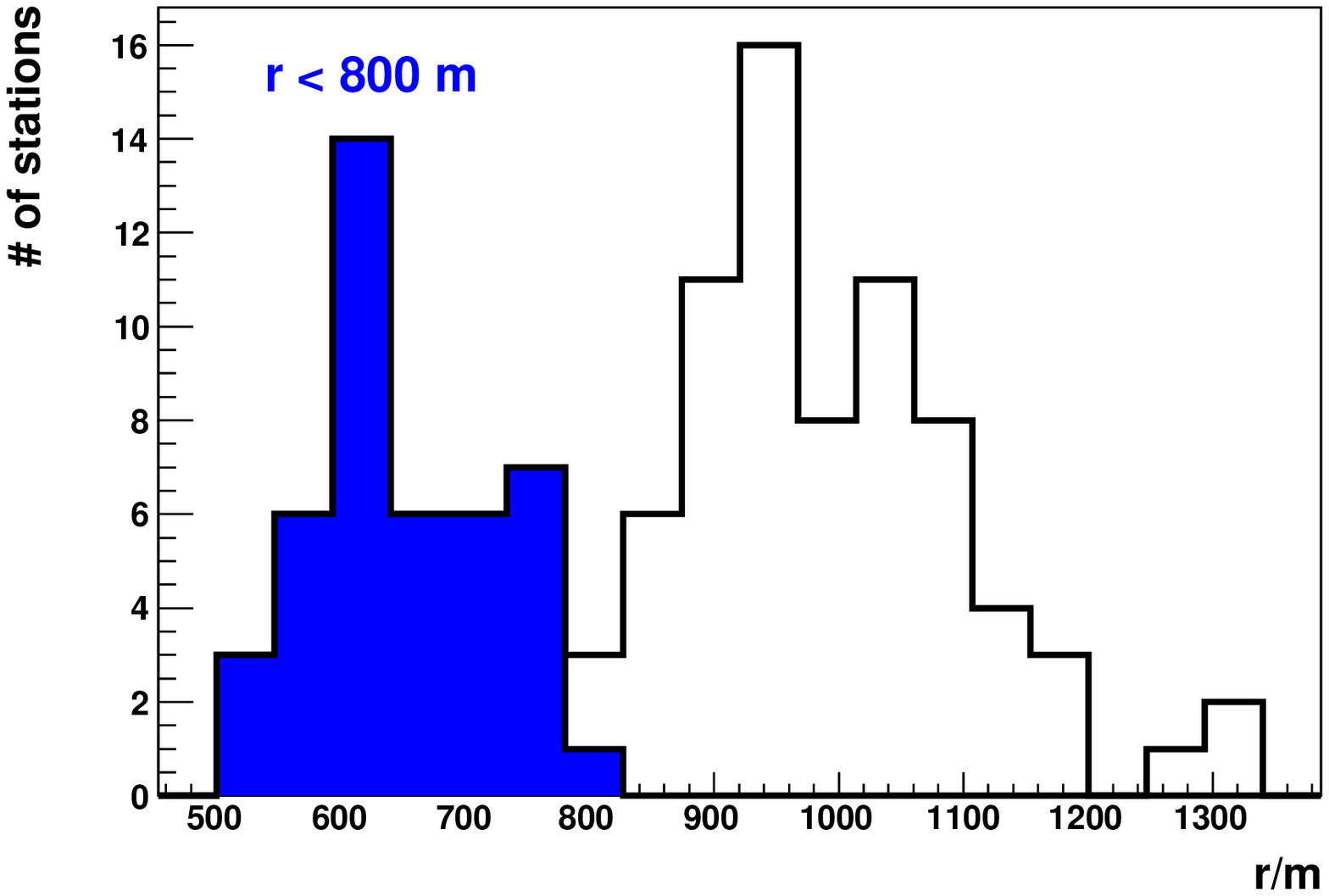}
  \end{center}
  \caption{\small Closest Station Distribution for \SI{e19.5}{eV} proton showers arriving at \SI{38}{\degree}. See text for details.}
  \label{fig:station_distribution}
  \end{minipage}
  \hspace{0.05\linewidth}
  \begin{minipage}[t]{0.47\linewidth}
  \begin{center}
    \includegraphics[width=\linewidth]{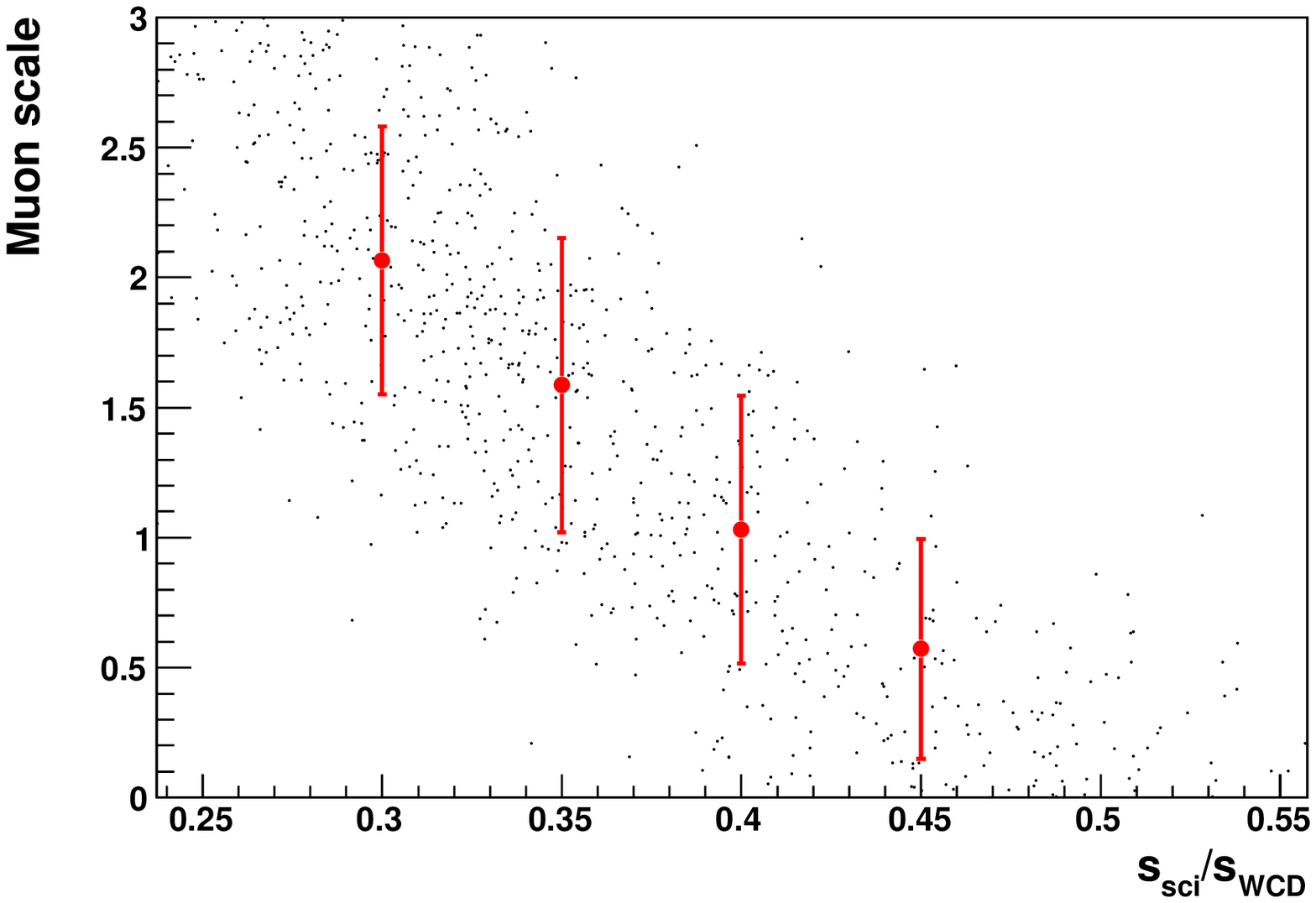}
  \end{center}
  \caption{\small Correlation between an artificial scaling of the muons in the shower and the ratio of the scintillator and WCD signals for the closest station.}
  \label{fig:signal_vs_mu_scale}
  \end{minipage}
\end{figure}

We are then interested in the fraction of events that will have a
station with a signal at a short distance to the axis. In Figure
\ref{fig:station_distribution} we show the radial distribution of the
non-saturated station closest to the axis for each event in a collection of
\SI{e19.5}{eV} protons arriving at a zenith angle of \SI{38}{\degree}.
One can see that 30\% of the events will have at least one
station within \SI{800}{m} of the axis (the region marked in blue). The
 peak around \SI{1000}{m} corresponds to events where the closest station is saturated. The radial distribution shows
two distinct peaks. The peak around \SI{950}{m} corresponds to events
where the WCD station closest to the axis is saturated and therefore
the station that enters the distribution is the closest WCD station
that is not saturated. If we consider the closest WCD station,
regardless of its saturated status, the fraction raises to 95\%.

The relation between the scintillator and WCD signals should provide,
in principle, a way to estimate the contribution from the muon
component, as shown in Figure \ref{fig:signal_vs_mu_scale}, where we
depict a collection of \SI{e19.5}{eV} protons arriving at
\SI{38}{\degree} with the vertical where we have scaled the number of
muons by an arbitrary factor between 0 and 3. We
use the linear correlation between the scintillator and WCD signals
for each component:
\begin{eqnarray}
s_{EM}^{\text{sci}} &=& \alpha_{EM} s_{EM}^{\text{WCD}} \label{eq:s_em_correlation}\\
s_{\mu}^{\text{sci}} &=& \alpha_{\mu} s_{\mu}^{\text{WCD}}
\end{eqnarray}
to determine $s_{\mu}^{\text{WCD}}$, the contribution from the muon
component to the WCD signal.

\begin{figure}[t]
  \begin{minipage}[t]{0.5\linewidth}
  \begin{center}
    \includegraphics[width=\linewidth]{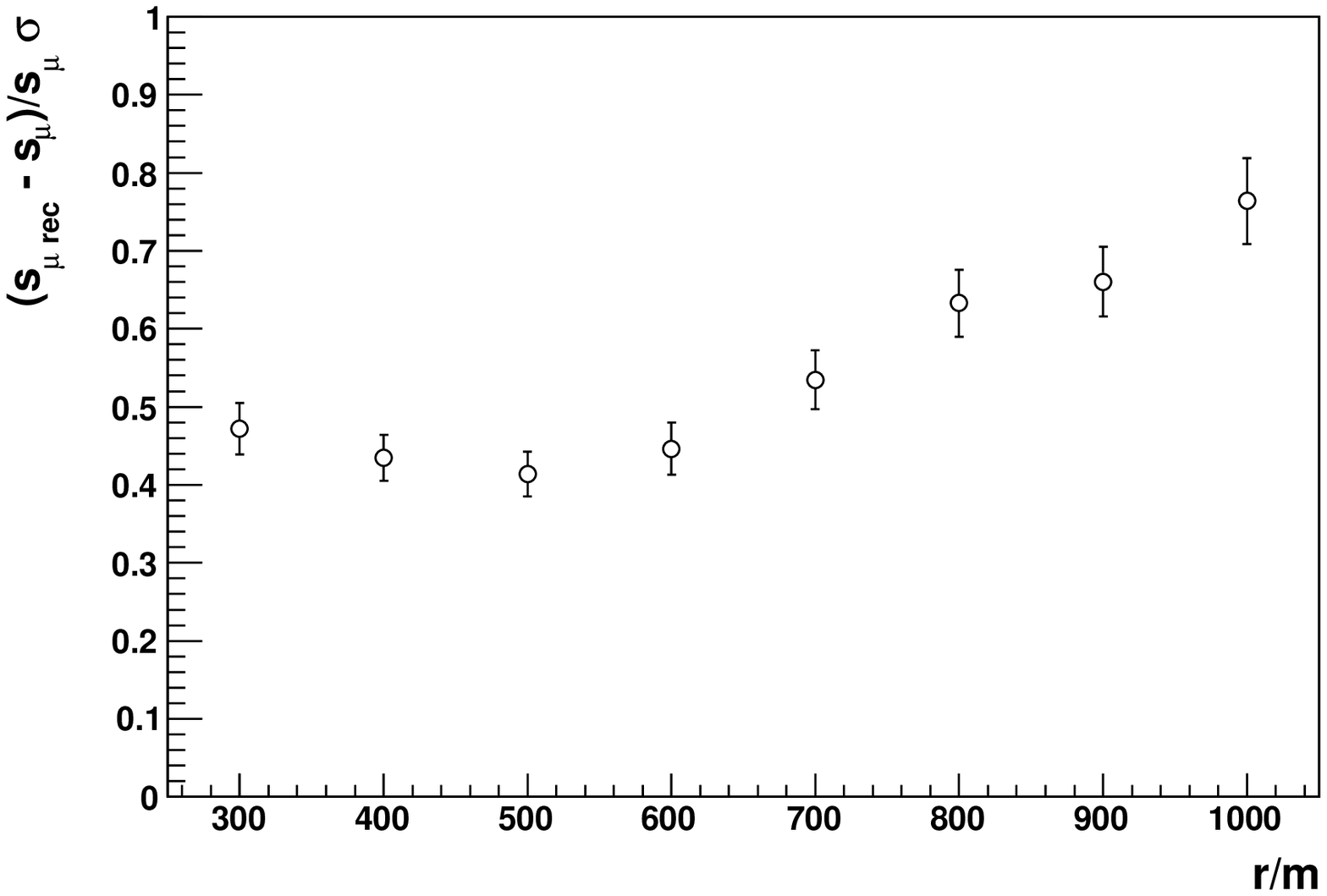}
  \end{center}
  \caption{\small Statistical uncertainty on $s_{\mu}$.}
  \label{fig:s_mu_delta_stat}
  \end{minipage}
  \begin{minipage}[t]{0.5\linewidth}
  \begin{center}
    \includegraphics[width=\linewidth]{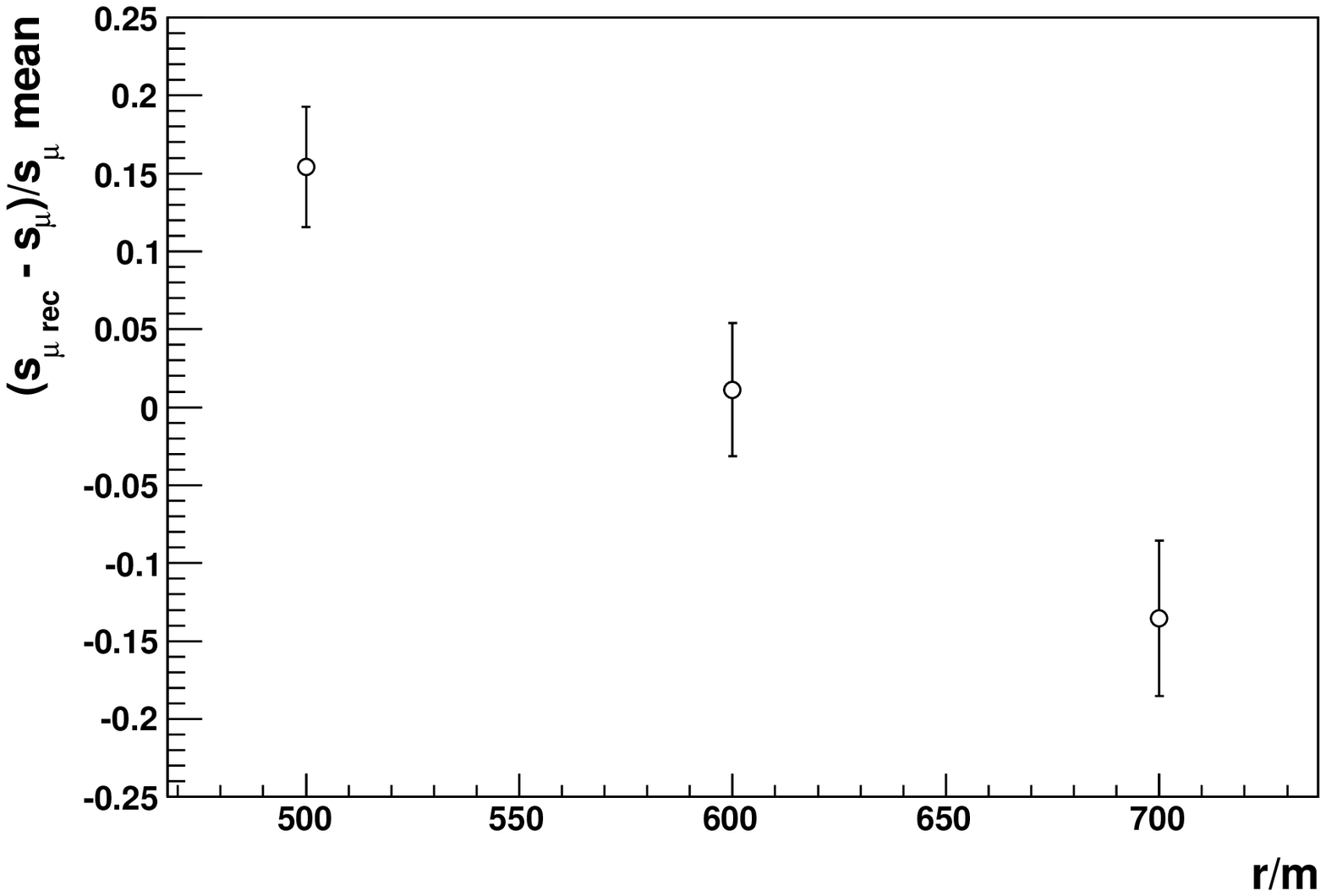}
  \end{center}
  \caption{\small Systematic deviations on $s_{\mu}$ arising from an incorrect distance to the core.}
  \label{fig:s_mu_delta_sys}
  \end{minipage}
\end{figure}

In Figure \ref{fig:s_mu_delta_stat} we show the statistical
uncertainty with which we can measure $s_{\mu}^{\text{WCD}}$ when we
use scintillator stations with an area of \SI{1.6}{m^2}. This
uncertainty is between 40\% and 50\% and could be reduced by
increasing the detector size. For \SI{10}{m^2} detectors it would be
around 30\%. This calculation was done for stations at fixed radii but
the uncertainty in the core position introduces another source of
uncertainty that is related to slight changes in $\alpha_{EM}$ in equation
\ref{eq:s_em_correlation}. The impact of this effect can be estimated
and is shown in Figure \ref{fig:s_mu_delta_sys}. From this we can
conclude that an uncertainty of \SI{30}{m} in the core position would
produce an uncertainty of less than 5\% in the reconstructed muon signal.

\section{Summary}

We have conducted a detailed study of the sensitivity of a combined
scintillator/WCD array to primary cosmic-ray mass
composition. Studying the response of this array to \SI{e18}{eV}
showers, we have concluded that adding photon converters on top of the
scintillators to enhance the signal from the electromagnetic component
of the shower does not increase the sensitivity. We have also
concluded that having many small scintillation detectors is better
than having few detectors of larger size since the optimum distance is
closer to the shower axis. Specifically, an array of \SI{3.2}{m^2}
detectors separated by \SI{375}{m} is better than an array of
\SI{9.7}{m^2} separated by \SI{750}{m}.

We have applied these ideas to the highest energies and determined
that, for \SI{e19.5}{eV} primaries, more than 30\% of the events will
have a pair of scintillator/WCD station within \SI{800}{m} of the
axis. Using this pair stations it would be possible to estimate the
signal from the muon component in the WCD tanks with an uncertainty of
50\% when the WCD station has an area of \SI{10}{m^2} and the
scintillator has an area of \SI{1.6}{m^2}. This uncertainty can be
reduced by increasing the scintillator size.


\clearpage

\end{document}